\documentstyle [12pt,twoside]{article}
\newread\epsffilein    
\newif\ifepsffileok    
\newif\ifepsfbbfound   
\newif\ifepsfverbose   
\newdimen\epsfxsize    
\newdimen\epsfysize    
\newdimen\epsftsize    
\newdimen\epsfrsize    
\newdimen\epsftmp      
\newdimen\pspoints     
\def\epsfbox#1{\global\def\epsfllx{72}\global\def\epsflly{72}%
   \global\def\epsfurx{540}\global\def\epsfury{720}%
   \def\lbracket{[}\def\testit{#1}\ifx\testit\lbracket
   \let\next=\epsfgetlitbb\else\let\next=\epsfnormal\fi\next{#1}}%
\def\epsfgetlitbb#1#2 #3 #4 #5]#6{\epsfgrab #2 #3 #4 #5 .\\%
   \epsfsetgraph{#6}}%
\def\epsfnormal#1{\epsfgetbb{#1}\epsfsetgraph{#1}}%
\def\epsfgetbb#1{%
\openin\epsffilein=#1
\ifeof\epsffilein\errmessage{I couldn't open #1, will ignore it}\else
   {\epsffileoktrue \chardef\other=12
    \def\do##1{\catcode`##1=\other}\dospecials \catcode`\ =10
    \loop
       \read\epsffilein to \epsffileline
       \ifeof\epsffilein\epsffileokfalse\else
          \expandafter\epsfaux\epsffileline:. \\%
       \fi
   \ifepsffileok\repeat
   \ifepsfbbfound\else
    \ifepsfverbose\message{No bounding box comment in #1; using defaults}\fi\fi
   }\closein\epsffilein\fi}%
\def\epsfsetgraph#1{%
   \epsfrsize=\epsfury\pspoints
   \advance\epsfrsize by-\epsflly\pspoints
   \epsftsize=\epsfurx\pspoints
   \advance\epsftsize by-\epsfllx\pspoints
   \epsfxsize\epsfsize\epsftsize\epsfrsize
   \ifnum\epsfxsize=0 \ifnum\epsfysize=0
      \epsfxsize=\epsftsize \epsfysize=\epsfrsize
     \else\epsftmp=\epsftsize \divide\epsftmp\epsfrsize
       \epsfxsize=\epsfysize \multiply\epsfxsize\epsftmp
       \multiply\epsftmp\epsfrsize \advance\epsftsize-\epsftmp
       \epsftmp=\epsfysize
       \loop \advance\epsftsize\epsftsize \divide\epsftmp 2
       \ifnum\epsftmp>0
          \ifnum\epsftsize<\epsfrsize\else
             \advance\epsftsize-\epsfrsize \advance\epsfxsize\epsftmp \fi
       \repeat
     \fi
   \else\epsftmp=\epsfrsize \divide\epsftmp\epsftsize
     \epsfysize=\epsfxsize \multiply\epsfysize\epsftmp   
     \multiply\epsftmp\epsftsize \advance\epsfrsize-\epsftmp
     \epsftmp=\epsfxsize
     \loop \advance\epsfrsize\epsfrsize \divide\epsftmp 2
     \ifnum\epsftmp>0
        \ifnum\epsfrsize<\epsftsize\else
           \advance\epsfrsize-\epsftsize \advance\epsfysize\epsftmp \fi
     \repeat     
   \fi
   \ifepsfverbose\message{#1: width=\the\epsfxsize, height=\the\epsfysize}\fi
   \epsftmp=10\epsfxsize \divide\epsftmp\pspoints
   \vbox to\epsfysize{\vfil\hbox to\epsfxsize{%
      \includegraphics{#1}%
      \hfil}}%
\epsfxsize=0pt\epsfysize=0pt}%

{\catcode`\%=12 \global\let\epsfpercent=
\long\def\epsfaux#1#2:#3\\{\ifx#1\epsfpercent
   \def\testit{#2}\ifx\testit\epsfbblit
      \epsfgrab #3 . . . \\%
      \epsffileokfalse
      \global\epsfbbfoundtrue
   \fi\else\ifx#1\par\else\epsffileokfalse\fi\fi}%
\def\epsfgrab #1 #2 #3 #4 #5\\{%
   \global\def\epsfllx{#1}\ifx\epsfllx\empty
      \epsfgrab #2 #3 #4 #5 .\\\else
   \global\def\epsflly{#2}%
   \global\def\epsfurx{#3}\global\def\epsfury{#4}\fi}%
\def\epsfsize#1#2{\epsfxsize}
\let\epsffile=\epsfbox

\begin{document}
\renewcommand{\thefootnote}{\alph{footnote}}
\par\noindent
to appear in: {\it Annals of the New York Academy of Sciences}
\vskip .1in
\par\noindent{\Large\bf Chaos in Cosmological Hamiltonians\footnote{
Some of the computations were facilitated by
computer time made available through the Research Computing Initiative at
the Northeast Regional Data Center (Florida) by arrangement with IBM.}}
\vskip .2in
\par\noindent
{HENRY E. KANDRUP\footnote{
Also with the Department of Physics and Institute for Fundamental 
Theory, University of Florida, Gainesville, Florida 32611.} AND JOHN DRURY}
\vskip .1in
\par\noindent{\it Department of Astronomy, University of Florida, Gainesville, 
Florida 32611}
\vskip .15in
\noindent{\small 
This paper summarises a numerical investigation which aimed to identify
and characterise regular and chaotic behaviour in time-dependent Hamiltonians
$H({\bf r},{\bf p},t)={\bf p}^{2}/2+V({\bf r},t)$, with 
$V=R(t)V_{0}({\bf r})$ or $V=V_{0}[R(t){\bf r}]$, where $V_{0}$ is a 
polynomial in $x$, $y$, and/or $z$ and $R(t){\;}{\propto}{\;}t^{p}$ is a 
time-dependent scale factor. When $p$ is not too negative, one can distinguish 
between regular and chaotic behaviour by determining whether an orbit segment 
exhibits a sensitive dependence on initial conditions. However, chaotic 
segments in these potentials differ from chaotic segments in time-independent 
potentials in that a small initial perturbation will usually exhibit a sub- or 
super-exponential growth in time. Although not periodic, regular segments 
typically exhibit simpler shapes, topologies, and Fourier spectra than do 
chaotic segments. This distinction between regular and chaotic behaviour is 
not absolute since a single orbit segment can seemingly change from regular to 
chaotic and visa versa. All these observed phenomena can be understood in 
terms of a simple theoretical model.}
\vskip .15in
\small
\centerline{\bf INTRODUCTION}
\vskip .1in
The past several decades have witnessed a growing recognition that chaotic
behaviour is seemingly ubiquitous in Nature, and that chaos could play an 
important role in many problems of astronomical interest, extending from 
stellar pulsations to galactic dynamics (see. e.g., Ref. [1]). It 
thus seems natural to consider the possibility that chaos could play 
an important role in problems related to cosmology, large scale structure, and 
quantum field theory in the early Universe. However, cosmological 
problems lead to an important new ingredient, namely the expansion of the 
Universe, which implies, e.g., that, even if the system of interest is 
symplectic, the Hamiltonian $H(t)$ generating the evolution will usually
have an explicit systematic time dependence. The obvious question then is: 
what effects, if any, will this time dependence have on the possibility of 
chaotic behaviour?

One example of some interest is the gravitational $N$-body problem,
as formulated for a large system of objects of comparable mass. In the 
context of isolated systems like individual galaxies, where the expansion of 
the Universe plays no role, the $N$-body problem is known to be chaotic in the 
sense that small initial perturbations in the locations of individual 
``particles'' grow exponentially (see. e.g., Refs. [2-4] and references 
contained therein). However, despite some preliminary investigations [5]
it is not yet clear whether this exponential instability persists for the
cosmological $N$-body problem, as formulated in a comoving frame that expands
with the Universe. Another cosmological example involving a 
time-dependent Hamiltonian is the problem of particle creation in the
early Universe, e.g., in the context of the phenomenon of preheating, which
leads to what Kofman, Linde \& Starobinsky [6] have termed a ``stochastic
resonance.''

But why might one expect that chaos will manifest itself differently in a
cosmological context than for systems characterised by a time-independent $H$?
For time-independent Hamiltonian systems, an orbit is usually said to be 
chaotic if and only if it has one or more positive Lyapunov exponents (see. 
e.g., Ref. [7]). However, these Lyapunov exponents can be 
defined in terms of the average properties of the stability matrix associated 
with an orbit, as evaluated along that orbit in an asymptotic $t\to\infty$ 
limit. Assuming canonical variables, the problem of stability, and 
hence the possibility of chaos, thus hinges on the properties of solutions to 
$${d{\delta}Z^{i}\over dt}={\cal J}^{ij}
{{\partial}^{2}H\over {\partial}Z^{j}{\partial}Z^{k}}{\Biggl |}_{Z_{0}}
{\delta}Z^{k}
{\;}{\equiv}{\;}{\Lambda}^{i}_{\;k}(t){\delta}Z^{k}, \eqno(1) $$
where ${\delta}Z^{i}$ denotes a perturbed phase space coordinate, 
${\cal J}^{ij}$ is the cosymplectic form [8], and 
${\Lambda}^{i}_{\;k}(t)$ is a function of $t$ because of its dependence on the 
unperturbed trajectory $Z^{\;i}_{0}(t)$. In particular, for a Hamiltonian 
$$H={1\over 2}{\bf p}^{2} + V({\bf r}), \eqno(2)$$
the configuration space perturbation satisfies
$${d{\delta}r^{a}\over dt^{2}}=
-{{\partial}^{2}V\over {\partial}r^{a}
{\partial}r^{b}} {\Biggr |}_{r_{0}(t)}\,{\delta}r^{b}.
\eqno(3) $$ 
If, e.g., the second derivative matrix 
${\partial}^{2}V/{\partial}r^{a}{\partial}r^{b}$ is constant and has at least
one negative eidenvalue, there exist solutions that grow exponentially
in time. If, however, that matrix acquires a secular time dependence this is
no longer guaranteed to be true. For example, the one-dimensional equation 
$${d^{2}{\delta r}\over dt^{2}}={\Omega}^{2}(t){\delta r}=
{{\Omega}_{0}^{2}\over t^{2}}\,{\delta r} \eqno(4)$$ 
admits solutions which exhibit a power law growth.
Just as an expanding Universe can convert an exponential Jeans instability 
into a milder power law instability [9], it might be expected to 
make chaotic orbits ``less chaotic.''

For time-independent Hamiltonian systems, sensitive dependence on initial
conditions and the existence of one or more positive Lyapunov exponents is
not the only way in which chaotic orbits differ from regular orbits. Regular
and chaotic orbits also have very different Fourier spectra. Because regular
orbits are multiply periodic, they will have computed spectra where (at least
if one integrates long enough) most of the power is concentrated at or near
a relatively small number of frequencies, whereas the spectra for chaotic
orbits should exhibit substantially broader band power.
(Strictly speaking, not every flow admitting one or more positive Lyapunov
exponents must have nonzero power for a continuous range of frequencies, but
one anticipates that, as a practical matter, positive 
Lyapunov exponents and broad band power go hand in hand [10].)

The situation is very different for Hamiltonian systems which manifest a
systematic secular time-dependence. In this case, one anticipates generically
that no orbit can be truly periodic, so that even spectra which one might
wish to interpret as corresponding to ``regular'' orbit segments should have 
Fourier spectra with broad band
power. This does not necessarily mean that an examination of the Fourier
spectra cannot be used to distinguish between regular and chaotic behaviour.
However, it {\it does} imply that any satisfactory discriminant based on an
inspection of Fourier spectra must be more subtle than determining whether
power is concentrated at or near a small number of frequencies.


The next section suggests a simple theoretical model based on a 
generalised Matthieu equation which can be used to make substantive 
predictions regarding the existence and manifestations of chaos in 
time-dependent Hamiltonian systems. This is followed by two sections which
describe in detail a collection of experiments which were performed to test 
the predictions based on this model and the results of those experiments. A 
final section  concludes by denumerating the principal conclusions.
\vskip .2in
\centerline{\bf THEORETICAL EXPECTATIONS}
\vskip .1in
To make reasonable predictions regarding the behaviour of orbits in 
time-dependent Hamiltonian systems, including possible sensitive dependence
on initial conditions, it is useful to understand precisely why chaotic
behaviour can arise in time-{\it in}dependent Hamiltonian systems. For
such systems, an exponentially sensitive dependence on initial conditions,
as manifested by the existence of one or more positive Lyapunov exponents,
is related to solutions to the linearised evolution equation (3) satisfied 
by a small initial ${\delta}r^{a}$, which can be interpreted as a 
time-dependent oscillator equation of the form
$${d^{2}{\delta}r^{a}\over dt^{2}}=-{\Omega}^{2}_{ab}(t){\delta}r^{b} .
\eqno(5)$$
If the matrix ${\Omega}^{2}_{ab}$ is constant in time and all its eigenvalues
are nonnegative, solutions to this equation involve stable oscillations, so 
that a small initial perturbation cannot grow exponentially. Alternatively, if
${\Omega}^{2}_{ab}$ is constant but has one or more negative eigenvalues, 
there {\it do} exist small perturbations that grow exponentially, which
implies a sensitive dependence on initial conditions. However, this latter
possibility does not seem very realistic: if there is always at least one
negative eigenvalue, the potential $V$ is not bounded from below!

The important point, therefore, is that instability is not necessarily
associated simply with the fact that the second derivative matrix
${\partial}^{2}V/{\partial}r^{a}{\partial}r^{b}$ has a negative eigenvalue.
Indeed, many nonintegrable potentials that admits large amounts of chaos, 
including all the finite order truncations of the Toda potential [11],
yield a second derivative matrix that is everywhere nonnegative. Rather, as 
has been discussed elsewhere [12-14] in the context of Maupertuis' Principle, 
where the flow associated with a time-independent $H$ is reinterpreted as a 
geodesic flow on a curved manifold, chaos in time-independent Hamiltonian 
systems can be understood as resulting from a parametric instability.

The idea is very simple. Given a knowledge of the unperturbed orbit,
${\bf r}_{0}(t)$, one could diagonalise eq.~(5) and then express the second
derivative matrix in terms of its Fourier transform to conclude that each
eigenvector ${\delta}r^{A}$ satisfies an equation of the form
$${d^{2}{\delta}r^{A}\over dt^{2}}=-{\Bigl[} {\cal C}^{A}_{0} + \sum_{\alpha}
{\cal C}^{A}_{\alpha}{\rm cos}\,({\omega}_{\alpha}t+{\varphi}_{\alpha})
{\Bigr]}{\delta}r^{A}, \eqno(6) $$
where, of course, the sum must be interpreted as a Stiltjes integral. 
The obvious point, then, is that, even if the coefficients 
${\cal C}^{A}_{\alpha}$ are sufficiently small that the term in brackets,
and hence the eigenvalues of the stability matrix, are always nonnegative,
there is the possibility of resonant behaviour leading to solutions that grow 
exponentially in time. One simple example, corresponding to the case where 
there is only one nonzero frequency ${\omega}_{\alpha}$, is the Matthieu 
equation [15], which can be written in the form
$${d^{2}{\xi}\over dt^{2}}=-(A + B {\rm cos}\,2t){\xi} .\eqno(7) $$

As is well known, a study of solutions to eq. (7) as a function of $A$ and $B$ 
reveals that the $A-B$ plane divides naturally into distinct, well defined 
regions corresponding to stable and unstable motions. In the stable regions, 
solutions to eq. (7) are purely oscillatory; in the unstable regions they 
exhibit a systematic exponential growth, i.e., 
$|{\xi}(t)|{\;}{\sim}{\;}{\rm exp}({\chi}t).$ 
The precise value of ${\chi}$ depends on $A$ and $B$, so that, e.g., unstable
values of $A$ and $B$ especially close to stable regions correspond to 
especially small (but still positive) values of ${\chi}$. However, the fact 
that ${\rm ln}\,|{\xi}|$ grows linearly in time is robust. 
Allowing for generalisations of eq. (7) which incorporate one or more
additional frequencies does not change the basic picture. In some cases,
${\xi}$ is bounded but, for other choices of parameters, ${\xi}$ grows in 
such fashion that ${\rm ln}\,|{\xi}|$ is reasonably well fit by a linear
growth law.

The obvious question then is: how do things change if the Hamiltonian $H$
acquires an explicit time dependence? Consider, e.g., the simplest possible 
time-dependence, where the potential is multiplied by an overall time-dependent
factor, so that
$$H={1\over 2}{\bf p}^{2} + R(t)V({\bf r}), \eqno(8)$$
with $R(t)$ a specified function of time. In this case, the natural
analogue of eq. (7) becomes
$${d^{2}{\xi}\over dt^{2}}=-R(t)(A + B {\rm cos}\,2t){\xi} ,\eqno(9) $$
(or, perhaps, a generalisation thereof with ${\rm cos}\,2t$ replaced by
${\rm cos}\,2{\tau}(t)$). 
Even if $R(t)$ evidence a systematic secular time-dependence, one can often
make reasonable distinctions between solutions to eq. (9) that do and do not
grow rapidly in time. However, in general the rapidly growing ``unstable'' 
solutions will not exhibit a purely exponential growth.

Consider, e.g., the case where $R(t)=R_{0}t^{p}$, with $p$ a real 
constant. Here trivial numerical computations reveal that, at least for values 
of $p$ somewhat larger than $p=-2$, the evolution of ${\xi}$ can be well 
understood in an adiabatic approximation. The factor 
$R{\;}{\propto}{\;}t^{p}$ in the potential implies that the instantaneous 
``natural'' frequencies ${\omega}$ with which ${\xi}$ grows or oscillates 
should scale as $R^{1/2}(t){\;}{\propto}{\;}t^{p/2}$; but, in 
the adiabatic approximation this leads to a time dependence
$${\int}\,dt\,{\omega}(t){\;}{\sim}{\;}{\int}\,dt\,R^{1/2}(t){\;}{\sim}{\;}
t^{1+p/2}. \eqno(10)$$
It follows that, for $p>0$, unstable solutions correspond to 
superexponential growth, so that 
${\rm ln}\,|{\xi}|{\;}{\sim}{\;}a+bt^{q},$
with $q=1+{p\over 2}>1$. Alternatively, for $-2<p<0$, unstable solutions 
correspond to subexponential growth with $q=1+{p\over 2}<1$. The adiabatic 
approximation fails for values of $p$ that are too small, the special case 
$p=-2$ corresponding instead to solutions that exhibit a (possibly 
oscillatory) power law time dependence.

The other obvious point is that a single initial condition evolved with eq.
(9) can exhibit transitions from stable to unstable motions and visa versa,
this corresponding to transitions between regular and chaotic behaviour. 
Solutions to the ordinary time-independent Matthieu equation involve either
stable or unstable motion, depending on the values of $A$ and $B$, which do
not change in time. However, incorporating a time dependence as in eq. (9) 
involves allowing for time-dependent ``dressed'' quantities 
${\hat A}=t^{p/2}A$ and ${\hat B}=t^{p/2}B$. In the adiabatic approximation, 
the time dependence involves ${\hat A}$ and ${\hat B}$ evolving through a 
sequence of values corresponding to a line in the $A-B$ plane. This line will 
in general intersect both stable and unstable regions, corresponding to 
intervals of both regular and chaotic motions. 

The basic inference is that, for Hamiltonian systems of the form given by 
eq.~(8) with $R{\;}{\propto}{\;}t^{p}$, power laws $p>0$ yield small 
perturbations of ``chaotic'' orbit segments that exhibit superexponential 
growth, whereas power laws $p<0$ yield small perturbations that exhibit 
subexponential or power law growth. For more complicated Hamiltonians, e.g., 
$H={1\over 2}{\bf p}^{2} + V[{\bf r}/R(t)],$
the simple scaling that leads to eq. (10) no longer holds. However, by analogy
with the preceeding one would anticipate that if the characteristic size of the
second derivative matrix ${\partial}^{2}V(t)/{\partial}r^{a}{\partial}r^{b}$
is increasing systematically in time, small perturbations of chaotic orbits 
should grow faster than exponentially, whereas the growth should be slower 
than exponential  if this matrix is decreasing systematically.
The examples described in the following sections corroborate this physical
expectation.
\vskip .2in
\centerline{\bf NUMERICAL EXPERIMENTS PERFORMED}
\vskip .1in
The numerical experiments described here were performed for time-dependent 
extensions of the time-independent potential 
$$V_{0}(x,y,z)=-(x^{2}+y^{2}+z^{2})+{1\over 4}(x^{2}+y^{2}+z^{2})^{2}-
{1\over 4}(ay^{2}z^{2}+bz^{2}x^{2}+cx^{2}y^{2}) ,\eqno(11) $$
which is itself an obvious three-dimensional generalisation of the 
two-dimensional dihedral potential of Armbruster, Guckenheimer, \& Kim [16]
for specific choices of parameter values. The simplest extension, most
easily compared with theory, involved introducing an overall multiplicative 
factor, setting 
$$V(x,y,z,t)=R(t)V_{0}(x,y,z) , \eqno(12) $$
with $R(t)=t^{p}$. Another alternative
involved mimicking the effects of comoving coordinates by setting
$$V(x,y,z,t)=V_{0}[R(t)x,R(t)y,R(t)z], \eqno(13) $$
again with $R(t)=t^{p}$.
Some computations focused on fully three-dimensional orbits. Others focused
on two-dimensional orbits with $z=p_{z}=0$. It was found that, at least in 
terms of their sensitive dependence on initial conditions, two- and 
three-dimensional orbits behaved very similarly but that, in terms of possible 
shapes, three-dimensional orbits exhibited a richer phenomenology.

Ensembles of ${\sim}{\;}1000$ initial conditions for use in two-dimensional 
simulations were generated 
by freezing the energy of the time-independent $H$ at a fixed value $E$,
setting $x=0$, uniformly sampling the energetically allowed regions of the
$y-p_{y}$ plane, and then solving for $p_{x}(x,y,p_{y},E)>0$. Initial
conditions for fully three-dimensional simulations were generated by freezing
the energy at $E$, setting $x=z=0$, uniformly sampling the allowed regions
of the $y-p_{y}-p_{z}$ cube, and solving for $p_{x}(x,y,z,p_{y},p_{z},E)>0$.
Each ensemble was evolved into the future for a time $t=256$ or longer,
with the initial time $t_{0}$ chosen to vary between $t_{0}=1.0$ and $t_{0}=
100$. A reasonably broad range of exponents $p$ was considered. The simulations
with the potential (12) allowed for $-1.5<p<1.5$. Those with the potential (13)
allowed for $-1<p<1$. 

The evolution equations were integrated using a fourth order Runge-Kutta
algorithm with fixed time step ${\delta}t$ ranging between $10^{-3}$ and
$10^{-4}$. The integrator solved simultaneously for the evolution of a small, 
linearised perturbation, renormalised at fixed intervals ${\Delta}t=1.0$, to 
obtain an estimate of the largest short time Lyapunov exponent (cf. Ref. [7]). 
When focusing on time-independent Hamiltonian systems, it is customary to 
record a running Lyapunov exponent that is a numerical approximation to the 
quantity
$${\chi}(t)=\lim_{{\delta}Z(0)\to 0}\,{1\over t}\;{\ln}\,{\Biggl[}
{||{\delta}Z(t)||\over ||{\delta}Z(0)||} {\Biggr]}, \eqno(14) $$
with $||\,.\,||$ the natural Euclidean norm, 
which converges towards the true Lyapunov exponent ${\chi}$ in a $t\to\infty$ 
limit. In the context of a time-dependent potential, it is more natural to 
record short time Lyapunov exponents (cf. Ref. [17])
${\chi}({\Delta}t_{i})$  for each interval ${\Delta}t$, which, for an
integration begun at time $t=0$, are related to ${\chi}(t)$ by [18]
$${\chi}({\Delta}t_{i})={{\chi}(t_{i}+{\Delta}t)(t_{i}+{\Delta}t)-
{\chi}(t_{i})t_{i}\over {\Delta}t} . \eqno(15) $$
Given such ${\chi}({\Delta}t_{i})$'s, the partial sums 
$${\xi}(t_{i})={1\over {\Delta}t}\;
\sum_{j=1}^{i-1}\,{\chi}({\Delta}t_{j})
={1\over {\Delta}t}\,  {\ln}\,{\Biggl[}
{||{\delta}Z(t_{i}+t_{0})||\over ||{\delta}Z(t_{0})|||} {\Biggr]} \eqno(16) $$
capture the net growth of the initial perturbation within a time $t_{i}$.

Plots of ${\chi}({\Delta}t_{i})$ and ${\xi}(t_{i})$ for individual orbit
segments were examined visually in an effort to identify clear distinctions 
between regular and chaotic behaviour. For those orbit segments deemed chaotic,
${\xi}(t_{i})$ was fitted to a growth law
$${\xi}=a+bt^{q} \eqno(17) $$
to determine (1) whether such a fit was reasonable and (2) whether the best 
fit yielded super- or sub-exponential growth.
Orbital data ${\bf r}(t)$ and ${\bf p}(t)$, and the associated Fourier spectra,
$|{\bf r}({\omega})|$ and $|{\bf p}({\omega})|$, were also inspected visually
in a search for distinguishing features. One aim was to determine whether
orbit segments deemed regular also had simpler topologies and/or simpler 
spectra than chaotic segments that manifested a sensitive dependence on 
initial conditions. The other was to search for evidence for abrupt 
transitions between chaotic and regular behaviour.
\vskip .2in
\centerline{\bf RESULTS OF THE EXPERIMENTS}
\vskip .1in
For values of $p$ that are not too negative, it is often possible to 
distinguish relatively clearly between regular segments, where 
${\chi}({\Delta}t_{i})$ fluctuates around zero, and chaotic segments, where,
if one averages over several time steps, ${\chi}({\Delta}t_{i})$ is 
usually larger than zero. This distinction becomes especially apparent if,
for an ensemble of segments in the same potential with the same value of $p$, 
one computes  $N[{\xi}(t_{fin})]$, the distribution of the final values
of ${\xi}$. This $N[{\xi}(t_{fin})]$ often corresponds to a bimodal 
distribution and, even when one seems to see only a single population, 
tracking the form of the distribution as a function of $p$ usually allows one 
to determine whether that population is regular or chaotic. 

This is illustrated in FIGURE 1, which was generated from an ensemble of 
$1000$ initial conditions with energy $E=1.0$ and $z=p_{z}=0$, evolved for the
interval $10.0<t<266.0$ in the potential $V=V_{0}[R(t){\bf r}]$ of eq. (13)
with $a$=1. The six panels correspond to different values of $p$ ranging
from $p=-0.6$ to $p=0.45$. It is clear that, for the time-independent case
with $p=0.0$, the distributions of ${\xi}$'s is bimodal, the peak near
${\xi}=0$ corresponding to regular segments, and the segments with larger
values of ${\xi}$ corresponding to chaotic orbits. (A longer time integration
reveals that the segments with $20<{\xi}<60$ correspond to ``sticky'' orbits
which, at early times, were trapped near regular islands by one or more 
cantori.) This bimodal behaviour persists for $p>0$, although the relative
abundance of ``regular'' segments increases rapidly with increasing $p$. 
Alternatively,  the relative abundance of regular segments decreases very
rapidly when $p$ becomes negative so that, for $p<0.1$ or so, a sample of 
$1000$ initial conditions is too few to yield an appreciable number of regular
segments. These changes in the relative abundance of regular and chaotic
segments probably reflect the specific form of the time-independent potential
$V_{0}({\bf r})$. For example, $p>0$ implies that the kinetic energy 
$K={\bf v}^{2}/2$ increases, the potential energy $V=t^{p}V_{0}(x,y)$ 
decreases in magnitude, and the total energy $E={\bf v}^{2}/2+t^{p}V_{0}(x,y)$
exhibits a modest systematic increase, with the net result that the orbits
tend to evolve in a regular or near-regular fashion in the ``trough'' of the
dihedral potential $V(t)$.

Another generic feature, also apparent in FIGURE 1, is that, even though 
increasing $p$ implies fewer chaotic segments, those segments that remain 
chaotic tend to be more unstable in the sense that the final ${\xi}(t_{fin})$ 
is larger. In part, this trend reflects the fact that, overall, the values of 
${\chi}({\Delta}t_{i})$ tend to be larger for larger values of $p$. However,
this trend also reflects the fact that, as expected, $p>0$ yields perturbations
that exhibit superexponential growth whereas $p<0$ yields subexponential 
growth. For fixed sets of initial conditions, this latter assertion was 
confirmed for each value of $p$ by (1) identifying a minimum value of ${\xi}$ 
that (seemingly) represents a sufficient criterion for chaotic behaviour, (2) 
fitting the computed ${\xi}(t_{i})$ for each chaotic segment to the power law 
(17), and (3) determining a mean slope for all the chaotic segments with given 
$p$. The results are exhibited in FIGURE 2 (a), where the error bars reflect 
the effects of reasonable variations in the value ${\xi}_{min}$ used to 
identify chaotic behaviour.

FIGURE 2 agrees with predictions in the sense that $p>0$ and $p<0$ yield, 
respectively, super- and sub-exponential growth. However, there 
is one new, not completely expected feature, namely that $q$ is not a monotonic
function of $p$. In particular, there appears to be a range of values of $p$,
say $-0.25<p<-0.15$ where, as probed by the aforementioned diagnostic, the
growth of a small initial perturbation is weaker than for both somewhat smaller
and somewhat larger values of $p$. Given this behaviour, it is especially
important to check whether the predicted behaviour for the simpler potential
$V=t^{p}V_{0}({\bf r})$ is confirmed by experiments. FIGURE 2 (b), which 
presents the analogue of FIGURE 2 (a) for the same set of initial conditions 
now evolved in the potential (12), indicates that, overall, the agreement 
between theory
and experiment is quite good, although some systematic differences are seen
for $p>0$. Note that the larger error bars for especially large and small
$p$ reflect the fact that, for these values of $p$, most of the orbits seem
regular or near-regular, with very small values of ${\chi}({\Delta}t_{i})$ and
${\xi}(t_{i})$.

A third general feature, also apparent from FIGURE 1, is that, for $p$'s 
somewhat larger than zero, a larger fraction of the segments have values of 
${\xi}$ well separated from both the low and high ${\xi}$ peaks than is the 
case for $p=0$. In most cases, these intermediate values appear to correspond 
to segments which change from chaotic to regular or, in some cases, from 
regular to chaotic. That this is the case is easily seen by computing either
${\chi}({\Delta}t_{i})$, which can exhibit abrupt systematic increases and 
decreases, or ${\xi}(t_{i})$, which can exhibit a nearly stepwise growth. Two 
examples of this behaviour are provided in FIGURE 3, both corresponding to
segments computed with $p=0.5$. The top two panels correspond to an
orbit segment which makes an abrupt transition from chaotic to regular
behaviour at $t{\;}{\sim}{\;}160$; the lower panels correspond to a segment
which exhibits a more erratic behaviour early on. It should also be evident
that, during the chaotic phases, ${\chi}({\Delta}t_{i})$ is evidencing a
systematic increase, so that ${\xi}(t_{i})$ grows faster than linearly in 
time, this corresponding to a perturbation that evolves superexponentially.

Inspection of individual orbit segments also reveals that segments which are
chaotic in the sense that they exhibit a sensitive dependence on initial
conditions tend to be manifestly more irregular in visual appearance. In
particular, regular segments typically have identifiable shapes and topologies 
which persist for relatively long periods of time, even as the orbital energy
changes by an order of magnitude or more. Pieces of two representative regular 
orbits evolved it the potential (13) with $a=1$ and, respectively, $p=0.3$ and 
$p=0.5$, are exhibited in the top four panels of FIGURES 4 and 5. Viewed over 
relatively short intervals ${\Delta}t<50$
or so, the first segment closely resembles a loop orbit in a time-independent 
potential. If, however, the orbit is tracked over longer intervals, one sees 
significant changes as the ``radius'' of the loop slowly decreases. The second 
segment, corresponding to the orbit used to generate FIGURE 3 (a) and (b),
exhibits more distinct variability than the loop orbit, but it is evident once 
again that the overall shape and topology are robust.

These regularities imply that, even though regular segments are not periodic,
their Fourier spectra are distinctly different, and simpler, than
the spectra for chaotic segments. For example, like true loop orbits in a
time-independent potential, regular segments that look loopy are characterised
by spectra $|x({\omega})|$ and $|y({\omega})|$ which are very similar in 
amplitude and shape. Moreover, in many cases the overall form of the spectrum 
can be interpreted as involving one or more peak frequencies ${\omega}$ whose
values exhibit a systematic drift over the course of time. This is particularly
evident in the final two panels of FIGURE 4, which exhibit $|x({\omega})|$ and 
$|y({\omega})|$ for the loopy regular orbit. At early times, when the orbit
rotates relatively slowly, the power for both $|x({\omega})|$ and 
$|y({\omega})|$ is concentrated at relatively low values of ${\omega}$ but,
as time elapses and the orbit begins to rotate more rapidly, power slides up
to high values of ${\omega}$. The composite spectra in FIGURES 4 (e) and (f) 
can be
understood, at least approximately, as representing the time integral of a set 
of narrow peaks which, as time elapses, move systematically towards higher 
frequency.
\vskip .2in
\centerline{\bf CONCLUSIONS}
\vskip .1in
This paper summarised a numerical investigation which focused on identifying 
meaningful definitions of regular and chaotic behaviour in time-dependent 
Hamiltonian systems of the type that one might expect to encounter in a 
cosmological setting. Especial attention focused on two- and three-dimensional 
Hamiltonians of the form $H({\bf r},{\bf p},t)={\bf p}^{2}/2+V({\bf r},t)$, 
with $V=R(t)V_{0}({\bf r})$ or $V=V_{0}[R(t){\bf r}]$, where $V_{0}$ is a 
polynomial in $x$, $y$, and $z$ and 
$R(t){\;}{\propto}{\;}t^{p}$ represents a time-dependent scale factor.
When $p$ is not too negative, one can distinguish between regular and chaotic 
behaviour by determining whether, over the time interval in question, an orbit 
segment exhibits a sensitive dependence on initial conditions. However, the
time-dependence of $H$ complicates the physics in several important ways. 
\par\noindent
1.~A single orbit can exhibit intermittent behaviour, changing from 
chaotic to regular and/or visa versa. 
\par\noindent
2.~A small perturbation of a chaotic segment will not in general exhibit an
average exponential growth. Rather, a computation of suitably defined short
time Lyapunov exponents shows that the phase space perturbation ${\delta}Z^{i}$
is often well fit by a growth law ${\rm ln}\,|{\delta}Z(t)|=a+bt^{q}$, with 
$q>1$ for $p>0$ and $q<1$ for $p<0$. An expanding Universe makes the effects 
of chaos milder; a contracting Universe makes them stronger. 
\par\noindent
3.~Regular segments are not periodic and, as such, do not have sharply peaked
Fourier spectra. However, the topology of regular segments {\it is} robust,
so that, e.g., a loop orbit continues to look loopy even as $|{\bf r}|$ and
$R(t)|{\bf r}|$ grow or shrink systematically. Moreover, the spectrum of a 
regular segment is simpler than that for a chaotic segment since, in many 
cases, the regular segment can be approximated as a sum of a few contributions 
of the form $Z(t){\;}{\sim}{\;}Z(0){\rm exp}[i{\Omega}(t)t]$, where 
${\Omega}(t)$ exhibits a simple secular variation.
\par\noindent
All these observed phenomena can be understood in terms of a simple theoretical
model based on a time-dependent generalisation of the Matthieu equation.
\vskip .2in
\centerline{\bf ACKNOWLEDGMENTS}
\vskip .1in
Work on this manuscript was completed while HEK was a visitor at the Aspen
Center for Physics, the hospitality of which is acknowledged gratefully.
\vfill\eject
\vskip .25in
\par\noindent
1. BUCHLER, J. R., S. L. GOTTESMAN, J. H. HUNTER \& H. E. KANDRUP, 1998. 
Nonlinear Dynamics and Chaos in Astrophysics. New York Academy of Sciences, 
New York, in press.
\par\noindent
2. KANDRUP, H. E. \& H. SMITH, 1991. Astrophys. J. {\bf 374}: 255.
\par\noindent
3. GOODMAN, J., D. HEGGIE \& P. HUT, 1993. Astrophys J. {\bf 415}: 715.
\par\noindent
4. KANDRUP, H. E., M. E. MAHON \& H. SMITH, 1994. Astrophys. J. {\bf 428}: 458.
\par\noindent
5. MELOTT, A. 1998. private communication.
\par\noindent
6. KOFMAN, L., A. LINDE \& A. A. STAROBINSKY, 1994. Phys. Rev. Lett. {\bf 73}:
3195.
\par\noindent
7. LICHTENBERG, A. J. \& M. A. LIEBERMAN, 1992. Regular and Chaotic Dynamics.
Springer-Verlag. Berlin. 
\par\noindent
8. ARNOLD, V. I. 1989. Mathematical Methods of Classical Mechanics. 
Springer-Verlag. Heidelberg.
\par\noindent
9. PEEBLES, P. J. E. 1993. Principles of Physical Cosmology. Princeton 
University Press. Princeton.
\par\noindent
10. TABOR, M. 1989. Chaos and Nonintegrability in Nonlinear Dynamics.
Wiley. New York.
\par\noindent
11. TODA, M. 1967. J. Phys. Soc. Japan {\bf 22}: 431.
\par\noindent
12. PETTINI, M. 1993. Phys. Rev. E {\bf 47}: 828.
\par\noindent
13. CERRUTI-SOLA, M. \& M. PETTINI, 1995. Phys. Rev. E {\bf 53}: 179.
\par\noindent
14. KANDRUP, H. E. 1997. Phys. Rev. E. {\bf 56}: 2722.
\par\noindent
15. WHITTAKER, E. T. \& G. H. WATSON, 1965. A Course of Modern Analysis.
Cambridge University Press. Cambridge.
\par\noindent
16. ARMBRUSTER, D., J. GUCKENHEIMER \& S. KIM, 1989. Phys. Rev. Lett. A {\bf
140}: 416.
\par\noindent
17. GRASSBERGER, P., R. BADII \& A. POLITI, 1988. J. Stat. Phys. {\bf 51}: 135.
\par\noindent
18. KANDRUP, H. E. \& M. E. MAHON, 1994. Astron. Astrophys. {\bf 290}: 762.
\vskip .2in
\vfill\eject
\vskip .1in
\par\noindent
FIGS. 1 -- The distribution $N[{\xi}(t_{fin})]$, with ${\xi}$ defined as in 
eq. (16), generated from an ensemble of initial conditions $E=1.0$ and
$z=p_{z}=0$ evolved for the interval $10<t<266$ in the potential (13) with
$a=1.0$ and variable $p$. (a) $p=-0.6$. (b) $p=-0.1$. (c) $p=-0.05$. (d)
$p=0.0$. (e) $p=0.2$. (f) $p=0.45$. 
\vskip .1in
\par\noindent
FIGS. 2 -- (a) The mean slope $q$ for chaotic segments generated as in FIG. 1, 
plotted as a function of $p$. (b) The analogue of (a) for chaotic segments
generated in the potential (12), again with $a=1.0$ and variable $p$.
\vskip .1in
\par\noindent
FIGS. 3 -- (a) and (b) The short time Lyapunov exponent 
${\chi}({\Delta}t_{i})$ and the cumulative ${\xi}(t_{i})$ computed for one
orbit with initial energy $E=1.0$ and $z=p_{z}=0$ evolved for the interval 
$10<t<266$ in the potential (13) with $p=0.5$. (c) and (d) The same quantities
for another orbit, again with $E=1.0$ and $z=p_{z}=0$, evolved for the same 
interval in the same potential. 
\vskip .1in
\par\noindent
FIGS. 4 -- Segments of a single trajectory with $E=1.0$ and $z=p_{z}=0$ 
evolved with $p=0.3$ in the potential (13) for the interval $10<t<266$, along
with the total power spectra, $|x({\omega})|$ and $|y({\omega})|$.
(a) $0<t<32$. (b) $64<t<96$. (c) $128<t<160$. (d) $192<t<224$. 
(e) $|x({\omega})|$. (f) $|y({\omega})|$.
\vskip .1in
\par\noindent
FIGS. 5 -- The analogue of FIG. 4 for another orbit with $E=1.0$ and 
$z=p_{z}=0$, now evolved with $p=0.5$.
\vfill\eject
\pagestyle{empty}
\begin{figure}[t]
\centering
\centerline{
        \epsfxsize=12cm
        \epsffile{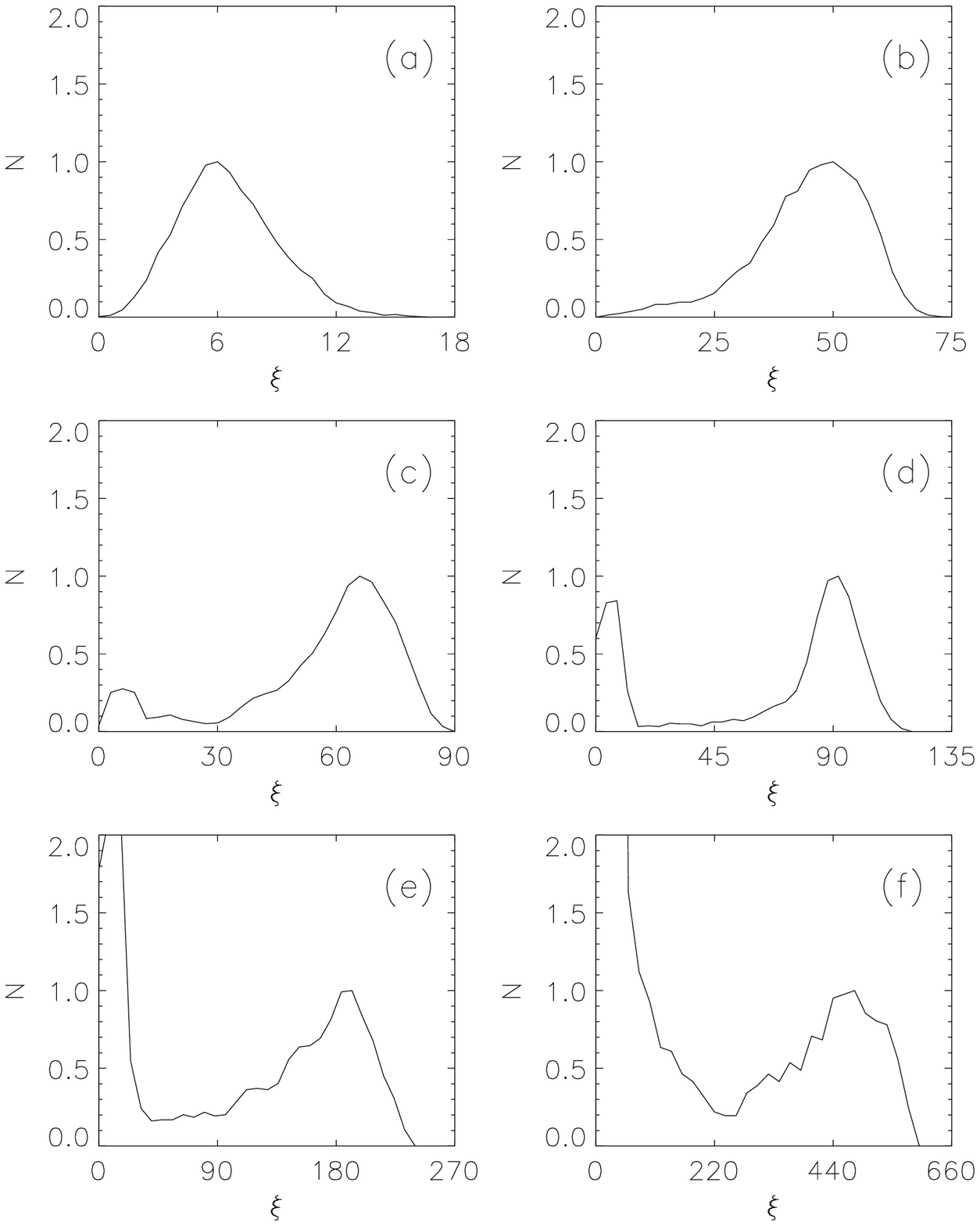}
           }
        \begin{minipage}{12cm}
        \end{minipage}
        \vskip -0.0in\hskip -0.0in
        \begin{center}\vskip .0in\hskip 0.5in
        Figure 1.
        \end{center}
\vspace{-0.2cm}
\end{figure}
\vfill\eject

\pagestyle{empty}
\begin{figure}[t]
\centering
\centerline{
        \epsfxsize=12cm
        \epsffile{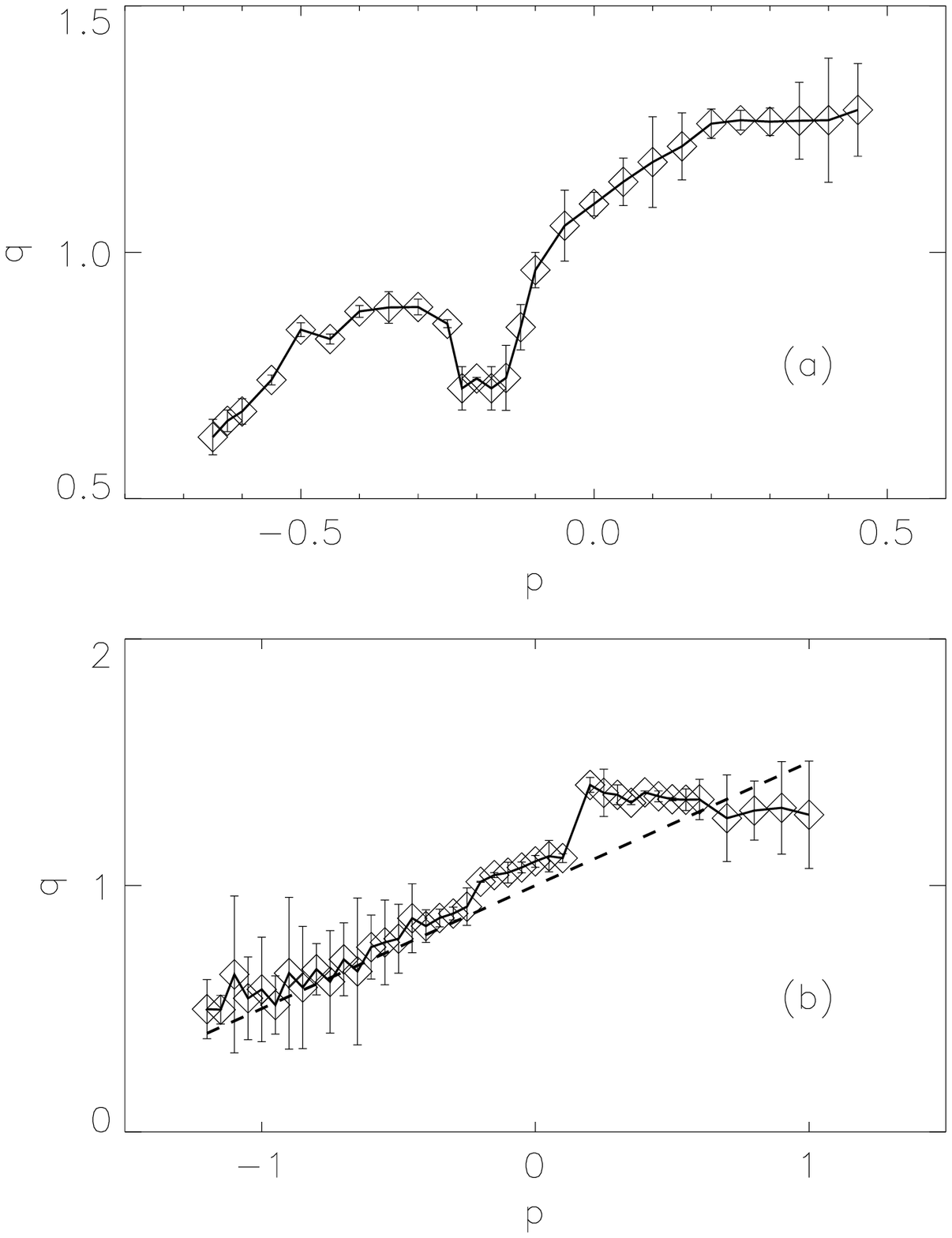}
           }
        \begin{minipage}{12cm}
        \end{minipage}
        \vskip -0.0in\hskip -0.0in
        \begin{center}\vskip .0in\hskip 0.5in
        Figure 2.
        \end{center}
\vspace{-0.2cm}
\end{figure}
\vfill\eject

\pagestyle{empty}
\begin{figure}[t]
\centering
\centerline{
        \epsfxsize=12cm
        \epsffile{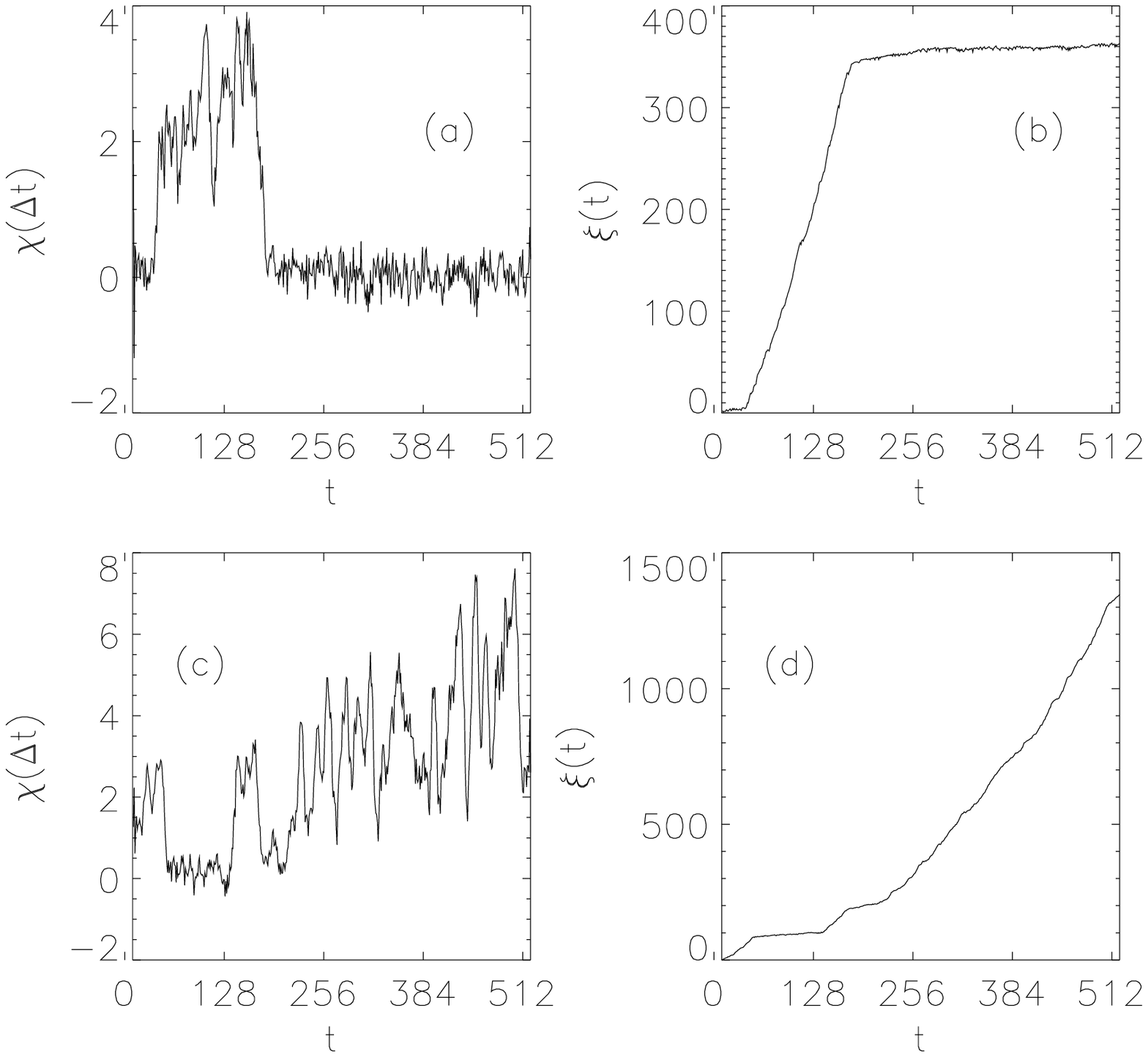}
           }
        \begin{minipage}{12cm}
        \end{minipage}
        \vskip -0.0in\hskip -0.0in
        \begin{center}\vskip .0in\hskip 0.5in
        Figure 3.
        \end{center}
\vspace{-0.2cm}
\end{figure}
\vfill\eject

\pagestyle{empty}
\begin{figure}[t]
\centering
\centerline{
        \epsfxsize=12cm
        \epsffile{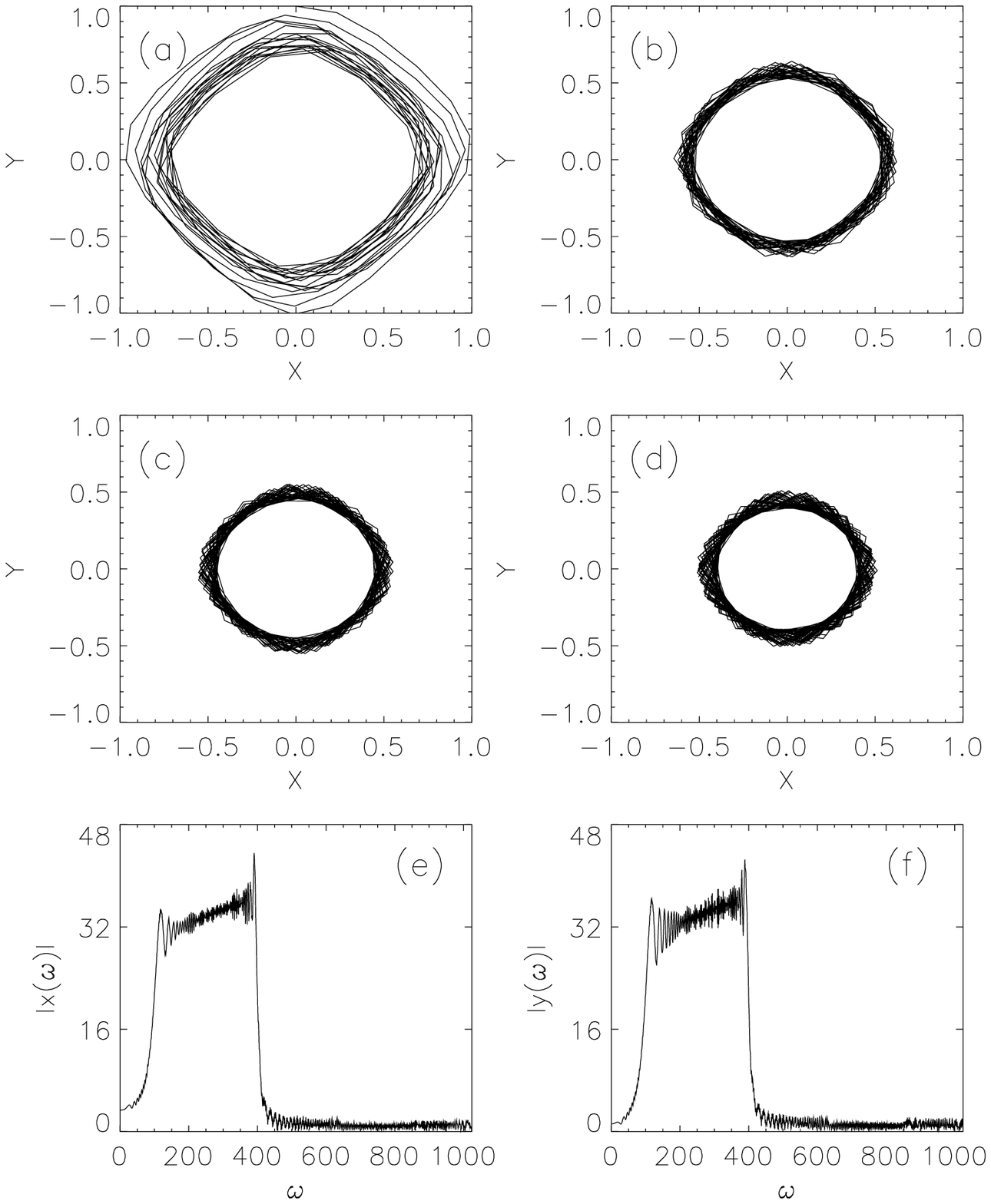}
           }
        \begin{minipage}{12cm}
        \end{minipage}
        \vskip -0.0in\hskip -0.0in
        \begin{center}\vskip .0in\hskip 0.5in
        Figure 4.
        \end{center}
\vspace{-0.2cm}
\end{figure}
\vfill\eject

\pagestyle{empty}
\begin{figure}[t]
\centering
\centerline{
        \epsfxsize=12cm
        \epsffile{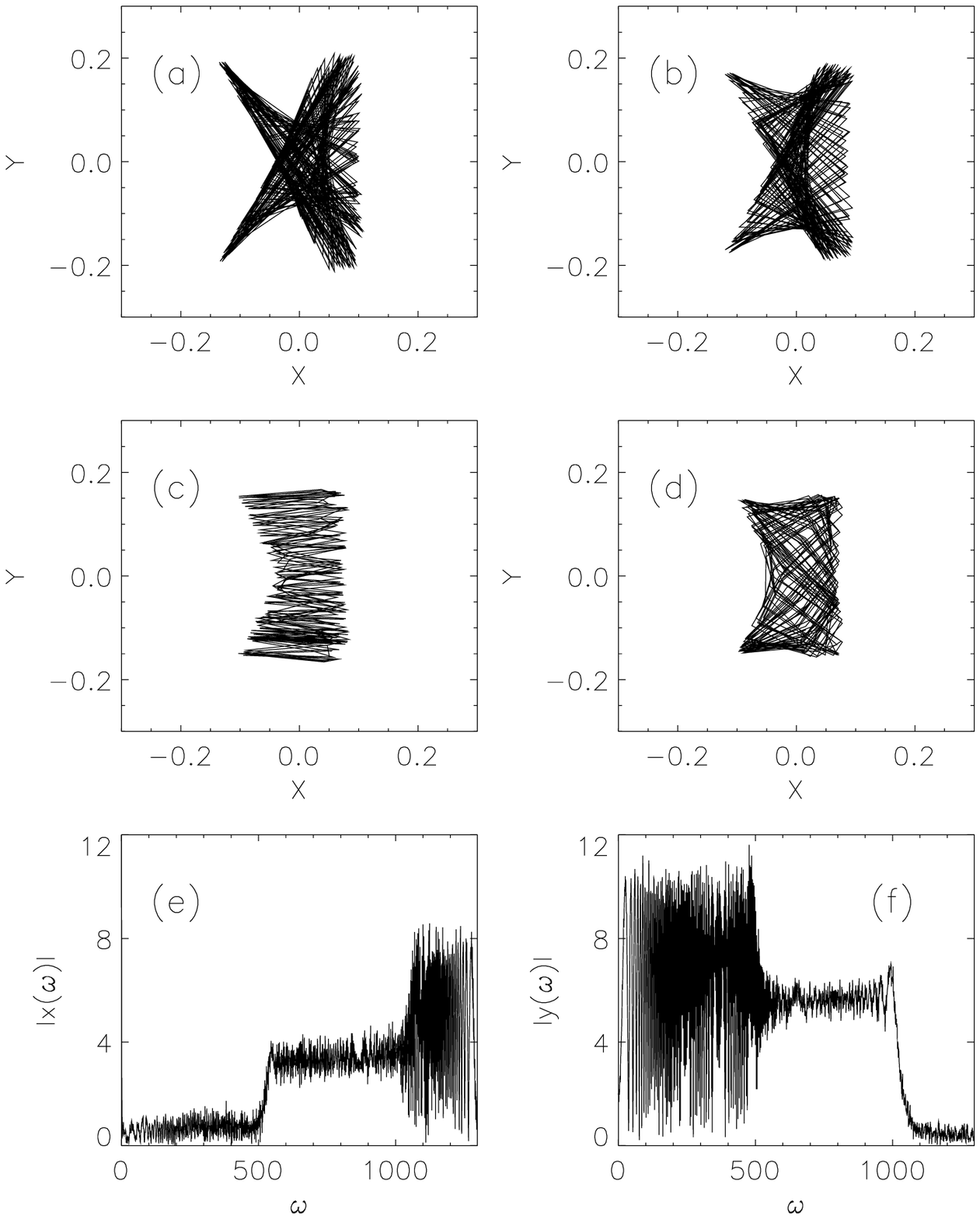}
           }
        \begin{minipage}{12cm}
        \end{minipage}
        \vskip -0.0in\hskip -0.0in
        \begin{center}\vskip .0in\hskip 0.5in
        Figure 5.
        \end{center}
\vspace{-0.2cm}
\end{figure}
\vfill\eject

\end{document}